\documentclass[prd,twocolumn,showpacs,nofootinbib]{revtex4}

\usepackage[usenames]{color}
\usepackage{amsfonts}
\usepackage{amsmath}
\usepackage{amssymb}
\usepackage{bm}
\usepackage{dcolumn}
\usepackage{epsfig}
\usepackage{graphicx}
\usepackage{graphics}
\usepackage[latin1]{inputenc}
\usepackage{latexsym}
\usepackage{rotating}
\usepackage{hyperref}
\usepackage{multirow}
\usepackage[caption=false]{subfig}

\newcommand{\<}{\begin{equation}}
\newcommand{\?}{\end{equation}}

\newcommand{\cC}{\mathcal{C}}

\DeclareMathOperator{\Real}{Re}

\newcommand{\fid}{\mathrm{fid}}
\newcommand{\surf}{\mathrm{surf}}

\newcommand{\out}{\mathrm{out}}

\newcommand{\Msolar}{\,M_{\odot}}

\newcommand{\circu}{s}

\begin{document}

\title{Gravitational wave constraints on the shape of neutron stars}

\author{Nathan~K.~Johnson-McDaniel}

 \affiliation{Theoretisch-Physikalisches Institut,
  Friedrich-Schiller-Universit{\"a}t,
  Max-Wien-Platz 1,
  07743 Jena, Germany}
  
\date{\today}

\begin{abstract}

We show that there is a direct relation between upper limits on (or potential future measurements of) the $m = 2$ 
quadrupole moments of slowly rotating neutron stars and the $l = m = 2$ deformation of the star's surface, in full general relativity,
to first order in the perturbation. This relation only depends on the star's structure through its mass and radius.
All one has to assume about the star's constituents is that the stress-energy tensor at its surface is that of a perfect fluid, which will be true with good accuracy in almost all the situations of interest, and that the magnetic field configuration there is force-free, which is likely to be a good approximation. We then apply this relation to the stars which have
direct LIGO/Virgo bounds on their  $m = 2$ quadrupole moment, below the spin-down limit, and compare with the expected surface deformation due to rotation. In particular, we find that LIGO observations have constrained the Crab pulsar's $l = m= 2$ surface deformation to be smaller than its $l = 2$, $m = 0$
deformation due to rotation, for all the causal equations of state we consider, a statement that
could not have been made just using the upper bounds on the $l = m = 2$ deformation from electromagnetic observations.

\end{abstract}


\maketitle

\section{Introduction}\label{Intro}

In the absence of significant internal stresses, objects bound by gravity are highly symmetric on large
scales. Indeed, Lindblom and Masood-ul-Alam~\cite{LMuA,Masood-ul-Alam} have shown that all static perfect fluid relativistic stars are spherical (with reasonable assumptions about the equation of state). And even if one
includes rotation, then equilibrium states of fluid stars must have a high degree of symmetry, as shown by Lindblom~\cite{Lindblom_PTRSA}: They must be axisymmetric, if one takes dissipation into account, and must
have reflection symmetry across the equator, at least in the Newtonian approximation (though this is conjectured to hold for relativistic stars, as well).

However, situations in which the internal stresses are nonnegligible are not uncommon. Indeed, it is obvious that many of the smaller (though still gravitationally bound) objects in the solar system have large-scale internal stresses that are comparable to their self-gravity. This is particularly true for asteroids and the smaller moons, which can be highly nonspherical (see, e.g.,~\cite{echo@JPL} and Fig.~2.2 in~\cite{Busch}), but is even true for the larger moons and the terrestrial planets, to a less immediately apparent degree. (For instance, see the model for
the Earth's gravitational field given in~\cite{PHKF}.) What is perhaps surprising is that such massive,
strong-gravity objects as neutron stars can also have large enough internal stresses to produce a nonnegligible large-scale deformation, with reasonable theoretical input. In particular, as first appreciated by Ruderman~\cite{Ruderman}, the star's solid crust can support a large-scale deformation, while an internal magnetic field will necessarily deform the star, as was first noted by Chandrasekhar and Fermi~\cite{CF}.

In most cases, these deformations will be quite small (maximum fractional deformations of $\sim 10^{-5}$, with considerably smaller deformations in most realistic cases). But there are certain scenarios in which the 
deformation could be (relatively) substantial ($\gtrsim 10^{-4}$, possibly as much as tens of percent in extreme cases): If there is a solid exotic phase (e.g., the
hadron-quark mixed phase) in the star's core, the entire star is solid (as is the case for crystalline superconducting quark stars),
or there is a large internal magnetic field of $\gtrsim 10^{15}$~G. (See, e.g.,~\cite{OwenPRL,J-MO2} for the solid case and~\cite{CFG,YKS,FR,CR} for the magnetic case.)

The mechanisms for creating large deformations on nonaccreting neutron stars are not clear, though it is worth pointing out that neutron stars are expected to be born with some deformation from the violent supernova explosion that forms them, and it is possible that this deformation may be frozen in to the solid parts, and only relax slowly over time, as discussed in~\cite{LIGO_psrs2010}. See also Sec.~2.2 in~\cite{Owen2009}
for a discussion of scenarios involving accretion. However, deformed neutron stars are an interesting enough prospect for gravitational wave detection that it is worthwhile to consider the more extreme cases, and see what bounds there are on large deformations. In most cases, the best bounds on the large-scale deformations of neutron stars come from electromagnetic
observations of their spin-down, which (as discussed in, e.g.,~\cite{LIGO_psrs2007}) constrain the star's $m = 2$ quadrupole moment, albeit with a
factor of $\sim 2$ uncertainty due to the star's unknown moment of inertia. There are also tighter, though less
direct, bounds that can be obtained from electromagnetic observations by folding in the star's braking index~\cite{Palomba}.

But the cleanest bounds come from gravitational wave observations, and a prominent accomplishment of the first generation of large-scale laser interferometer gravitational wave detectors has been setting direct upper bounds on the $m = 2$ quadrupole moment of certain known
neutron stars, below the bounds that are given by electromagnetic observations of the spin-down, or indirect
limits given by the star's age~\cite{LIGO_Crab, LIGO_psrs2010, LIGO_CasA, LIGO_Vela}. See~\cite{Owen2009, Pitkin, Astone} for recent
reviews. It is customary to
quote these upper bounds in terms of a fiducial ellipticity for the star to give a feeling for the size of the
deformation. However, this fiducial ellipticity only gives a direct measure of the star's deformation in the uniform density, Newtonian limit, and it is \emph{a priori} unclear how to relate this to a measure of the deformation in more realistic cases. It is our purpose here to show how one can
convert such fiducial ellipticities into a more physically meaningful, fully relativistic measure of the shape of the star's
surface. This relativistic shape measure gives a more tangible interpretation of present upper bounds on the
gravitational radiation emitted from known pulsars than does the fiducial ellipticity. In particular, it allows one to compare the bounds on the star's $l = m = 2$ deformation directly with various other deformations
(e.g., the star's deformation due to rotation, or the deformations of other astronomical objects). [\emph{Nota bene} (N.B.): While there is also the possibility of an $m = 1$ quadrupole deformation, as discussed by Jones~\cite{Jones}, we will not
consider this situation here, since there are no final results for searches for such radiation---though see~\cite{Gill} for
preliminary results. Similarly, we do not consider radiation from
higher multipole deformations, as such radiation is suppressed by factors of the star's rotational speed over the speed of light compared to the
$l = 2$ radiation, and thus not searched for.]

The relativistic shape measure we define is inspired by the measure of the horizon deformation of a tidally distorted
black hole used by Taylor and Poisson~\cite{TP} and involves the ratio of the star's longest polar circumference to its equatorial circumference. (Similar ratios have also been used in
numerical studies of distorted black holes~\cite{Anninosetal,BS, BCST}.) The conversion from fiducial ellipticity to
relativistic shape measure
only depends upon the star's structure through its mass and radius (with reasonable assumptions about the magnetic field and
shear stresses at the star's surface). This simple conversion comes from the simple relation between the
metric on the star's perturbed surface and the surface value of the time-time component of the
Regge-Wheeler gauge~\cite{RW}
metric perturbation (given by Damour and Nagar~\cite{DN}), and the relationship between the surface value of this component and the
amplitude of the star's quadrupole gravitational radiation, given in~\cite{J-MO2} (the relation with the quadrupole moment was first given by Hinderer~\cite{Hinderer}). (As discussed in~\cite{J-MO2}, these relations assume slow rotation, using the results of Ipser~\cite{Ipser} and Thorne~\cite{Thorne} to show that it is appropriate to calculate the gravitational waves from the rotating star using the quadrupole moment calculated in the static limit. This is 
quite a good approximation for the stars we consider, which are rotating well below the Kepler limit---at most
$\sim 10\%$ of it.)

One can compute the deformation of the star due to rotation in the same manner, using Hartle's classic calculation~\cite{Hartle67} of the metric of a slowly
rotating relativistic star. While there is no reason to assume any sort of correlation between a star's $l = m = 2$
deformation and its rotational deformation,\footnote{Indeed, millisecond pulsars, which have the largest rotational deformations, for a fixed mass, have some of the tightest spin-down constraints on their $m = 2$ quadrupoles---see, e.g.,
Table~1 in~\cite{LIGO_psrs2010}; one can obtain a star's spin-down ellipticity by dividing the ellipticity bound given in the table by the value of $h_0^{95\%}/h_0^\mathrm{sd}$.} the rotational deformation gives a convenient scale against which
to compare the bounds on the $l = m = 2$ deformation. We then give these conversions (from fiducial
ellipticity and rotational frequency to surface deformation) for a variety of equations of state (EOSs), including
the case of a strange quark star (for which one needs to modify the calculations slightly). We also compare 
the bounds on the $l = m = 2$ surface deformation with the surface deformation due to rotation for the four
stars with known spin periods for which the LIGO/Virgo bounds on the $l = m = 2$ deformation are near
or below the (fiducial) spin-down limit. (We also note that models of strange quark stars with large radii can give significantly larger moments of inertia than the maximum of $3\times10^{45}\text{ g cm}^2$ assumed in the
LIGO/Virgo papers, which increases the uncertainty on the spin-down limit.)
Additionally, we show that these results will receive negligible corrections from non-perfect fluid contributions
in the cases of interest (with the possible exception of corrections to the force-free nature of the magnetic
field configuration, which could be large enough to affect the results).

The paper is structured as follows: We define the shape measure, give some intuition about its properties, and detail how to compute it from
the fiducial ellipticity in Sec.~\ref{e22}. We then compute the surface deformation due to rotation in Sec.~\ref{e20_comp}, and compare the sizes of these two deformations for a variety of equations of state and pulsars of interest in Sec.~\ref{comp}. Finally, we conclude and consider the outlook for future bounds on the
$l = m = 2$ deformations of known neutron stars in Sec.~\ref{concl}. In the Appendix, we show that the corrections to these results due to the non-perfect fluid nature of real neutron stars are likely negligible in realistic cases. We use geometrized units throughout (i.e., take $G = c = 1$, where $G$ is Newton's
gravitational constant and $c$ is the speed of light).

\section{The relativistic shape measure}\label{e22}

\subsection{Definition}\label{def}

First, recall that the fiducial ellipticity is defined by
\<
\epsilon_{\text{fid, } 22} := \frac{I_{xx} - I_{yy}}{I_{zz}^\fid},
\?
where $I_{xx}$ and $I_{yy}$ are the star's \emph{actual} moments of inertia about the axes other than its rotation axis (the difference between the two is related to the star's $m = 2$ quadrupole moment, which is the observable quantity) and $I_{zz}^\fid = 10^{45}\text{ g cm}^2$ is the \emph{fiducial} moment of inertia of the
star about its rotational axis. [As is discussed in, e.g., Sec.~VI~B of~\cite{LIGO_psrs2007}, the actual moment of inertia of neutron stars is expected
to lie in the range ($1$--$3$)$\times 10^{45}\text{ g cm}^2$, on theoretical grounds, but there are no
measurements of this quantity for any neutron star. And as we shall see, one can obtain even larger moments of
inertia, up to at least $\sim 5.5\times10^{45}\text{ g cm}^2$, for models of strange quark stars with large radii.]

In contrast, our relativistic measure of the deformation will be constructed solely from quantities on the star's surface, which could be measured (in principle) by a physical observer, viz., the star's equatorial circumference, $\circu_{\text{eq, } 22}$, and its longest polar circumference, $\circu_\text{pol, 22, max}$, giving
\<\label{e_surf_def}
\epsilon_{\text{surf, } 22} := \frac{\circu_\text{pol, 22, max} - \circu_{\text{eq, } 22}}{\circu_{\text{eq, } 22}}
\?
(cf.\ the calculation of the surface deformation of a tidally deformed black hole in Sec.~VIII of Taylor and Poisson~\cite{TP}, and the calculations of the shape of deformed black holes in~\cite{Anninosetal,BS, BCST}).\footnote{Note that in the
rotational case, these circumferences can no longer be determined directly from timelike or null geodesics, due to frame dragging.}
One can compute the quantities entering this definition using the metric on the star's
surface, so all we need to do is write this metric in terms of the star's quadrupole moment (along with its
unperturbed mass and radius). 

To gain some intuition about the properties of this deformation measure, we consider a Newtonian
ellipsoid, with semi-axes $(1+\eta)R$, $(1-\eta)R$, and $R$, for which we have $\epsilon_{\text{surf, } 22} = \eta/2$ (to first order in $\eta$). [Here we have used the expression $2\pi a[1 - e^2/4 + O(e^4)]$ for the
circumference of an ellipse with semi-major and semi-minor axes $a$ and $b$ and eccentricity
$e = \sqrt{1 - b^2/a^2}$.] If the ellipsoid has uniform density, then $I_{xx} - I_{yy} = 4\eta I_{zz} = (8/5)\eta MR^2$ [note that here we are using the \emph{true} moment of inertia about the $z$-axis, $I_{zz} = (2/5)MR^2$, not the fiducial one, here and still working to first order in $\eta$], so its Newtonian ellipticity is $\epsilon_{N\text{, } 22} = 4\eta$, giving
$\epsilon_{\text{surf, } 22} = \epsilon_{N\text{, } 22}/8$.

\subsection{Computation}

Let us now relate the relativistic shape measure to the star's fiducial ellipticity. The star's surface metric is given in standard spherical coordinates by Eqs.~(85) and~(86) in Damour and Nagar~\cite{DN}
\<
ds^2 = R^2\left(1 + 2\frac{\delta R}{R}Y_{lm}\right)(d\theta^2 + \sin^2\theta d\phi^2),
\?
where $R$ is the star's undeformed radius, $Y_{lm}$ is a spherical harmonic (we shall only consider $l =  m = 2$ in our discussion), and the deformation is given by [Eqs.~(90) and (92) in Damour and Nagar~\cite{DN}~]
\<\label{dRR_orig}
\frac{\delta R}{R} = -\frac{1}{2}H_0(R)\biggl[\frac{2}{\cC} - 1 +  \frac{\cC}{2} + \frac{\cC^2/4}{1-\cC}
+ \frac{\cC}{4}\frac{d\log H_0(R)}{d\log R}\biggr].
\?
Here $H_0(R)$ is the surface value of the time-time component of the $l = m = 2$ piece of the Regge-Wheeler gauge~\cite{RW} perturbation of the star's metric [see Eq.~(24) in~\cite{J-MO2} for the full perturbed metric, which comes from the stellar perturbation formalism of Thorne and Campolattaro~\cite{TC}~] and $\cC := 2M/R$ is the star's compactness, where $M$ is the star's mass. (Note that Damour and Nagar denote our $H_0$ by merely $H$
and define a compactness of
half our $\cC$, which they denote by $c$.) This expression assumes a nonrotating star, but this
is quite a good approximation for the stars in question, since they are rotating slowly enough that one can also treat their rotation as a perturbation of a static star, and the two perturbations are independent to first order (since their multipolar structure is different). The corrections due to rotation will thus be at most $\sim 1\%$. Additionally, as discussed in~\cite{J-MO2}, the gravitational quadrupole radiation
from a slowly rotating star can be computed from the static quadrupole moment of the star with no rotation, as shown by Ipser~\cite{Ipser}
and Thorne~\cite{Thorne}.

We now want to relate this to the $m = 2$ quadrupole moment of the deformed star. Here we consider the
quadrupole moment amplitude $Q_{22}$ given by
\<
\label{Q22}
Q_{22} = \int_0^\infty\delta\rho(r)r^4dr
\?
in the Newtonian limit (the relativistic version is read off of the asymptotic expansion of the metric). (Here $\delta\rho$ is the $l = m = 2$ piece of the star's density perturbation.) We thus substitute the expression for $H_0(R)$ in terms of $Q_{22}$ from Eqs.~(39) and~(41) in~\cite{J-MO2}, giving
\<\label{dRR}
\begin{split}
\frac{\delta R}{R} &= -\frac{1}{2}\frac{\pi Q_{22}}{M^3}\bar{H}_0(R)\biggl[\frac{2}{\cC} - 1
+ \frac{\cC}{2} + \frac{\cC^2/4}{1-\cC}\\
&\quad + \frac{\cC}{4}\frac{d\log\bar{H}_0(R)}{d\log R}\biggr]\\
&=: -\frac{1}{2}\frac{\pi Q_{22}}{M^3}\bar{f}_{22}(\cC).
\end{split}
\?
Here we have defined $\bar{H}_0(R)$ to stand for $H_0(R)$ with unit amplitude (i.e., $c_1\to 1$), so
\<
\begin{split}
\bar{H}_0(R) &= \left(\frac{2}{\cC} - 1\right)\frac{\cC^2/2 + 3\cC - 3}{1 - \cC} + \frac{6}{\cC}\left(1 - \frac{1}{\cC}\right)\\
&\quad \times \log\left(1 - \cC\right),
\end{split}
\?
and have also defined the ``correction factor'' $\bar{f}_{22}(\cC)$. [Since $\bar{H}_0(R)$ only depends on $\cC$, the correction factor indeed only
depends on $\cC$. Also note that $d\log\bar{H}_0(R)/d\log R = -d\log\bar{H}_0(R)/d\log\cC$.] 

It is necessary to make a small change to the computation of $\delta R/R$ to treat the case of strange quark stars, where the
density does not go to zero at the surface. This affects the present calculation because the pressure perturbation no
longer goes to zero at the surface, so one must add on $\pi H_0(R)R^2\rho_-$ to the expression for $\delta R/R$ given in Eq.~\eqref{dRR_orig}, where $\rho_-$ is the density just inside the star's surface. This addition comes from noting that [Eqs.~(88) and (91) in Damour and Nagar~\cite{DN}, recalling that their $K$ is the negative of ours]
\<\label{dRRH0K}
\frac{\delta R}{R} = \left.\left[\left(1 - \frac{1}{\cC}\right)H_0 + \frac{K}{2}\right]\right|_{r=R},
\?
so we need to take into account corrections to the relation between $K$ and $H_0$ from the nonzero surface density. Such corrections can be read off from Eq.~(39a) in Ipser~\cite{Ipser}, which gives an addition to
$K$ of $-4\pi R^2\delta p$ at the surface, where $\delta p$ is the $l = m = 2$ piece of the Eulerian pressure perturbation, and is denoted $-p_1$
by Ipser. We then use the stress-energy conservation expression for $\delta p$ given in Eq.~(35)
of~\cite{J-MO2} to obtain $\delta p = -H_0(R)\rho_-/2$ at the surface, which then yields the expression given above.
[N.B.:\ We neglect the shear stress terms in all of these relations, which we shall show is a good
approximation in the Appendix.] However, this contribution only decreases the
star's deformation by $\lesssim 5\%$ in the case we consider. 

We now compute the desired arclengths, finding (to first order in $\delta R/R$)
\<
\circu_{\text{eq, } 22} = R\int_0^{2\pi}\biggl[1 + \sqrt{\frac{15}{32\pi}}\frac{\delta R}{R}\cos2\phi\biggr]d\phi = 2\pi R
\?
and
\<
\begin{split}
\circu_{\text{pol, } 22} &= 2R\int_0^{\pi}\biggl[1 + \sqrt{\frac{15}{32\pi}}\frac{\delta R}{R}\sin^2\theta\cos 2\phi_0\biggr]d\theta\\
&= 2\pi R\biggl[1 + \sqrt{\frac{15}{128\pi}}\frac{\delta R}{R}\cos 2\phi_0\biggr],
\end{split}
\?
where we have used $\Real Y_{22} = \sqrt{15/32\pi}\sin^2\theta\cos2\phi$ and considered the longitude line
with azimuthal angle $\phi_0$. We can now write $\epsilon_{\text{surf, } 22}$ in terms of $Q_{22}$, or the fiducial $l = m = 2$ ellipticity $\epsilon_{\text{fid, } 22} = \sqrt{8\pi/15}Q_{22}/I_{zz}^\fid$, using
Eqs.~\eqref{e_surf_def} and~\eqref{dRR}, yielding
\<
\begin{split}
\epsilon_{\text{surf, } 22} &= \sqrt{\frac{15\pi}{512}}\frac{Q_{22}}{M^3}\bar{f}_{22}(\cC) = \frac{15}{64}\frac{I_{zz}^\fid}{M^3}\bar{f}_{22}(\cC)\epsilon_{\text{fid, } 22}\\
&=: f_{22}(M,\cC)\epsilon_{\text{fid, } 22}.
\end{split}
\?
Here $f_{22}(M,\cC)$ gives the conversion factor between the ($l = m = 2$) fiducial ellipticity and the $l = m = 2$ surface
deformation. [N.B.: Since $\bar{f}_{22}(\cC) > 0$, we have $\delta R/R < 0$, and thus the maximum value of
$\circu_{\text{pol, } 22}$ is given by taking $\cos2\phi_0 = -1$.]

Note that in the Newtonian limit, we have $\epsilon_{\text{surf, } 22} = (3/8)(I_{zz}^\fid/MR^2)\epsilon_{\text{fid,  } 22}$. 
This does not agree with the Newtonian calculation for the uniform density ellipsoid given at the end of Sec.~\ref{def}, which gives a coefficient of $5/16$, but that is to be expected, since the present computation assumes a fluid star, whose surface will deform to be an equipotential of its perturbed gravitational field, while the ellipsoid was taken to be rigid.

\section{Calculation of the rotational deformation}\label{e20_comp}

It is interesting to compare upper bounds on an $l = m = 2$ deformation to the $l = 2$,
$m = 0$ deformation induced by the star's rotation. Here, one can use the slow-rotation
results of Hartle~\cite{Hartle67} to perform the calculation. We start by recalling that [Eq.~(86) in Damour and Nagar~\cite{DN}; cf.\ Eq.~(25a) in Hartle and Thorne~\cite{HT}~]
\<\label{dRR2}
\frac{\delta R}{R} = \frac{\delta r}{r} + \frac{1}{2}K(R),
\?
where $\delta r/r$ gives the fractional position of the star's deformed surface, and $K(R)$ is the ($l = 2$, $m = 0$
piece of the) angular
component of the metric perturbation, evaluated at the star's surface.\footnote{We neglect the $l = 0$ change to the star's radius, since
it gives a second-order correction to the calculation of the relativistic shape measure.} [Here, following Hartle, we take the
angular dependence to be just
the Legendre polynomial portion of $Y_{20}$, viz., $(3\cos^2\theta - 1)/2$, without the normalization factor of $\sqrt{5/4\pi}$ present in
the spherical harmonic. Additionally, note that we have reversed the sign of Damour and Nagar's $K$ to
correspond to the conventions of Hartle~\cite{Hartle67} and~\cite{J-MO2}.]

We will obtain expressions for the quantities entering Eq.~\eqref{dRR2} in terms of the surface values of
Hartle's $h_2$ and $k_2$, the $l = 2$, $m = 0$ components of the time-time and angular metric perturbations [see Eqs.~(66)--(69) in Hartle~\cite{Hartle67}], noting that  $K = 2k_2$ [cf.\ Hartle's Eq.~(66) and Eq.~(24) in~\cite{J-MO2}]. We will then solve the equations Hartle gives for $h_2$ and $v:= h_2 + k_2$ to obtain these
surface values for a given stellar model.

Specifically, to obtain $\delta r/r$, we note
that Eq.~(146) in Hartle~\cite{Hartle67} gives
an expression for Hartle's $\xi_2/a$, which is the same as our $\delta r/r$, so we have
\<
\frac{\delta r}{r} = \left(2 - \frac{2}{\cC}\right)h_2(R) - \frac{8M^2}{3\cC^3}\left(\Omega - \frac{J\cC^3}{4M^3}\right)^2,
\?
where
\<
\begin{split}\label{h_2(R)}
h_2(R) &= A\left[- \frac{6(1-\cC)}{\cC^2}\log(1 - \cC) -\frac{6}{\cC} + 3 + \cC + \frac{\cC^2/2}{1-\cC} \right]\\
&\quad + \frac{J^2\cC^3}{8M^4}\left(1 + \frac{\cC}{2}\right)
\end{split}
\?
[Eqs.~(137) and~(139) in Hartle~\cite{Hartle67}~],
$\Omega$ is the magnitude of the star's angular velocity, and $J = I_{zz}\Omega$ is the magnitude of its
angular momentum, where $I_{zz}$ is the star's (true) moment of inertia. Here we have noted that Hartle's $\zeta := 2/\cC - 1$. We obtain $K(R)$ from Eqs.~(137) and (139)--(141) in Hartle~\cite{Hartle67}, giving
\<\label{K(R)}
\begin{split}
\frac{1}{2}K(R) &= A\left[\frac{6}{\cC} + 3 - \cC +3\left(\frac{2}{\cC^2} - 1\right)\log(1 - \cC)\right]\\
&\quad - \frac{J^2\cC^3}{8M^4}\left(1 + \cC\right).
\end{split}
\?

The amplitude $A$ is given by
solving [Eqs.~(125) and~(126) in Hartle~\cite{Hartle67}~]
\begin{subequations}
\label{rot-eqs}
\begin{align}
v' &= -2\phi'h_2 + \frac{1}{6}\left(1 + r\phi'\right)[r^3j^2(\bar{\omega}')^2 - 2r^2(j^2)'\bar{\omega}^2],\\
h_2' &= \left\{-2\phi'+\frac{r}{[r-2m(r)]\phi'}\left[4\pi(\rho + p) - \frac{2m(r)}{r^3}\right]\right\}h_2 \nonumber\\
&\quad - \frac{2v}{r[r-2m(r)]\phi'} + \frac{1}{6}\left\{r\phi' - \frac{1}{2[r-2m(r)]\phi'}\right\}
\nonumber\\
&\quad \times r^3j^2(\bar{\omega}')^2 - \frac{1}{3}\left\{r\phi' + \frac{1}{2[r-2m(r)]\phi'}\right\}r^2(j^2)'\bar{\omega}^2,
\end{align}
\end{subequations}
where primes denote derivatives with respect to the (Schwarzschild) radial coordinate $r$; $\rho$ and $p$
are the star's energy density and pressure, respectively; we have [cf.\ Eqs.~(26)--(29) and (40) in Hartle~\cite{Hartle67}~]
\begin{subequations}
\label{rot-eqs-defs}
\begin{align}
\phi' &= \frac{m(r) + 4\pi r^3 p}{r[r - 2m(r)]},\\
\phi(R) &= \log(1 - 2M/R)/2,\\
m(r) &:= 4\pi\int_0^r\rho(\bar{r})\bar{r}^2d\bar{r},\\
j &:= \left[1 - \frac{2m(r)}{r}\right]^{1/2}e^{-\phi};
\end{align}
\end{subequations}
and the frame-drag parameter $\bar{\omega}$ is given by [Eq.~(46) in Hartle~\cite{Hartle67}~]
\<
\label{omega-bar}
\frac{1}{r^4}(r^4j\bar{\omega}')' + \frac{4}{r}j'\bar{\omega} = 0.
\?
[Note that Hartle denotes $2\phi$ by $\nu$ and $2\phi'$ by $\nu_R$. We have chosen to use the same
notation as in~\cite{J-MO2} to avoid confusion with our use of $\nu$ for the star's spin frequency here.
Similarly, we use $\rho$ and $p$ for the star's energy density and pressure instead of Hartle's $E$ and $P$.
Finally, we generally suppress the arguments of functions when we are not evaluating them at a
specific point, unless we feel that the argument needs to be included for clarity, as with $m(r)$.]

We slightly streamline the solution process for $v$ and $h_2$ given above Hartle's Eq.~(146):  We still write the solution as a linear combination of the solutions to the homogeneous and inhomogeneous equations and determine the unknown
coefficients by matching the solution and its first derivative to the known exterior
solutions at the surface of the star [Eqs.~\eqref{h_2(R)} and~\eqref{K(R)}, recalling that $v = h_2 + K/2$]. However, we find that it is not necessary to use Hartle's more involved inner boundary
conditions, and that we can simply integrate the inhomogeneous
equations starting from values of $0$ at $r_0$, the inner radius where
we impose our inner boundary condition when solving the enthalpy version of the Oppenheimer-Volkov equations~\cite{Lindblom}---see the
discussion at the end of Sec.~III in~\cite{J-MO2}. We also take boundary conditions for the homogeneous
equations [i.e., Eqs.~\eqref{rot-eqs} with the source terms that do not contain $v$ and $h_2$ omitted] of $h_2(r_0) = r_0^2/R^2$ and $v_2(r_0) = -2\pi(p_c + \rho_c/3)r_0^4/R^2$ [cf.\ Eqs.~(128) and
(144) in Hartle~\cite{Hartle67}~]. Here $p_c$ and $\rho_c$ denote the central values of the pressure and
energy density, which are the same as those at $r_0$, in our treatment. The boundary condition for $\bar{\omega}$ is given by $\bar{\omega}(0) = \text{const}$; one then scales the final result to give the desired angular velocity, using the known solution of $\bar{\omega}(r) = \Omega - 2J/r^3$ outside the star [Eq.~(47) in Hartle~\cite{Hartle67}~], noting that the surface deformation scales as $\Omega^2$.

Additionally, we
convert Eqs.~\eqref{rot-eqs} and~\eqref{omega-bar} to enthalpy form (i.e., with the enthalpy $h$ as the dependent variable; despite notation, this has no relation to $h_2$). This was first done for the frame-drag equation~\eqref{omega-bar} in Sec.~4.1 of K.~Lockitch's thesis~\cite{Lockitch}. [The transformed frame-drag equation also appears in a slightly different form in Appendix~A of~\cite{Readetal}. Since this transformation is simply performed by dividing through by $h'(r)$ and substituting $r\to r(h)$, we choose not to show the transformed
equations explicitly.] Finally, we perform the matching at the surface using $h_2$ and its first derivative to obtain the amplitude $A$ [cf.\ Eq.~\eqref{h_2(R)}], and use the $v_2$ matching as a check. (This check indicates that our calculations are
accurate to better than $\sim 2\%$, usually much better.)

We must also consider the changes that need to be made to this calculation to treat strange quark stars, with their
nonzero surface density. Due to Hartle's method of locating the surface of the rotating star using a first integral to
the equations of hydrostatic equilibrium, no change is necessary in that portion of the calculation. However, we do need to modify the surface matching used to
obtain the amplitude. The calculation of the moment of inertia remains unchanged, since only the second
derivative of $\bar{\omega}$ is discontinuous at the surface. But the contribution to
$(j^2)'$ from the density evaluated just inside the star's surface is $-8\pi R\rho_-/(1 -\cC)$ [from Eqs.~\eqref{rot-eqs-defs}, noting that $e^{-2\phi(R)} = 1/(1 - \cC)$~].
Thus, the matching of the solutions to Eqs.~\eqref{rot-eqs} at the surface needs to be adjusted using the replacements
\begin{subequations}
\begin{align}
v'_\out &\to v'_\out + \frac{4\pi}{3}\left(2 + \frac{\cC}{1-\cC}\right)\frac{R^3\rho_-\bar{\omega}^2(R)}{1 - \cC},\\
\nonumber
h'_{2,\text{ out}} &\to h'_{2,\text{ out}} + \frac{4\pi R^2\rho_- h_2(R)}{M} + \frac{4\pi}{3}\left(\frac{2}{\cC} + \frac{\cC}{1-\cC}\right)\\
&\quad\times\frac{R^3\rho_-\bar{\omega}^2(R)}{1 - \cC},
\end{align}
\end{subequations}
where the subscript ``out'' denotes the solution outside the star.

One can compare the values for the quadrupole obtained using the Hartle slow-rotation expansion with the fits to a
fully relativistic calculation for a $1.4\Msolar$ star given in Eq.~(73) and Table~5 in
Frieben and Rezzolla~\cite{FR}. Here one uses Eq.~(26) in~\cite{HT} to obtain the
quadrupole from the slow-rotation calculation (the expression given in~\cite{Hartle67} is in error). We find quite close agreement for the
four EOSs we both consider---at worst, $\sim 2\%$ for the BBB2 and GNH3 EOSs, and at best, better than $0.1\%$, for
the APR and SLy EOSs. (What we call the SLy EOS is the same as Frieben and Rezzolla's SLy4; we give
further discussion of these EOSs in Sec.~\ref{comp}.) Our
calculations of the moment of inertia all agree to better than $0.1\%$.

Now doing the calculations of the equatorial and polar circumferences, we have,
as previously,
\<
\begin{split}
\circu_{\text{eq, } 20} &= R\int_0^{2\pi}\left[1 - \frac{1}{2}\left(\frac{\delta R}{R}\right)_{20}\right]d\phi\\
&= 2\pi R\left[1 - \frac{1}{2}\left(\frac{\delta R}{R}\right)_{20}\right],
\end{split}
\?
\<
\begin{split}
\circu_{\text{pol, } 20} &= 2R\int_0^{\pi}\left[1 + \frac{1}{2}\left(\frac{\delta R}{R}\right)_{20}(3\cos^2\theta - 1)\right]d\theta\\
&= 2\pi R\left[1 + \frac{1}{4}\left(\frac{\delta R}{R}\right)_{20}\right],
\end{split}
\?
where $(\delta R/R)_{20}$ is calculated by combining together Eqs.~\eqref{dRR2}--\eqref{K(R)}.
Thus, we have
\<
\begin{split}
\epsilon_{\text{surf, } 20} &:= \frac{\circu_{\text{eq, } 20} - \circu_{\text{pol, } 20}}{\circu_{\text{eq, } 20}} = -\frac{3}{4}\left(\frac{\delta R}{R}\right)_{20}\\
&=: g_{20}(M,\cC)\nu^2
\end{split}
\?
to first order in $\delta R/R$, where we have used the known scaling of the surface deformation with $\Omega^2$ in the final equality to write $\epsilon_{\text{surf, } 20}$ in terms of the star's spin frequency, $\nu$, defining the conversion
factor $g_{20}(M,\cC)$.
We have also introduced
a minus sign, compared with the $l = m = 2$ version, so that $\epsilon_{\text{surf, } 20}$ will be positive.

For the purposes of comparison, we relate our $\epsilon_{\text{surf, } 20}$
to the coordinate radius ellipticity or flattening in the Euclidean limit (and to first order in the deformation). This coordinate radius ellipticity is defined by 
\<
\epsilon_c = \frac{r_e - r_p}{r_e},
\?
where $r_e$ and $r_p$ are the star's equatorial and polar radii, respectively, so we have $\epsilon_c = 2\epsilon_{\text{surf, } 20}$ in the given limit. (Here we have noted that the eccentricity of an
ellipse is related to the flattening by $e^2 = 2\epsilon_c - \epsilon_c^2$. Note also that Frieben and Rezzolla~\cite{FR} define their surface deformation to be $r_e/r_p-1$ [see their Eq.~(68)], though this agrees with $\epsilon_c$
to first order in the deformation.)

\section{Results and discussion}\label{comp}

\begin{figure*}[htb]
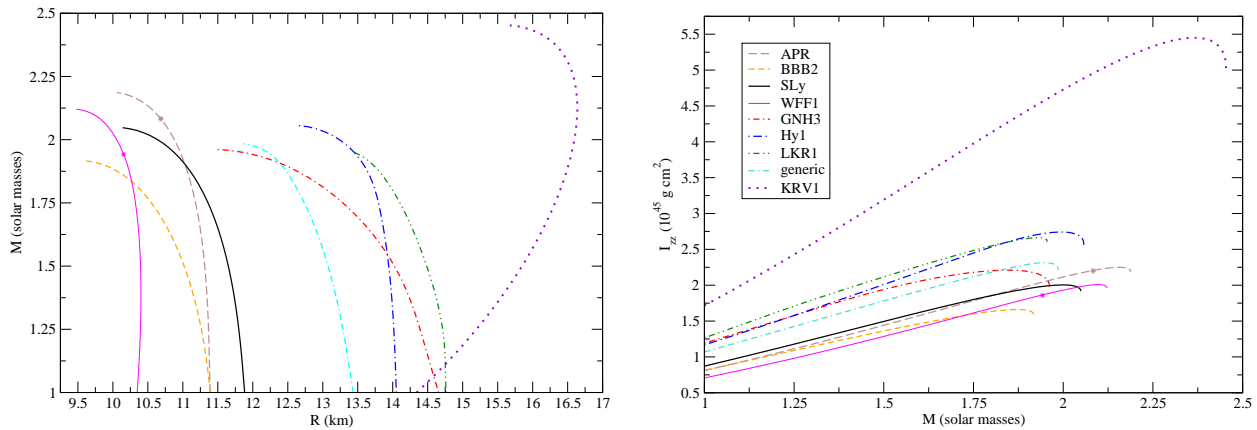

\centering
\subfloat{
\epsfig{file=M_vs_R_EOSs_shape_paper_WFFs.eps,width=8cm,clip=true}
}
\quad
\subfloat{
\epsfig{file=Izz_vs_M_EOSs_WFF.eps,width=8cm,clip=true}
}
\caption{\label{MvsR} The mass-radius relation  \emph{(left)} and moment of inertia versus mass \emph{(right)} for the EOSs considered in this paper.
For the APR and WFF
EOSs, the asterisks mark the maximum mass for which the stars do not contain any acausal matter (i.e., for which
the central density is smaller than the density at which the EOS has a sound speed greater than the speed of light).}
\end{figure*}

Here we show the results of our calculations for a representative selection of EOSs, whose mass-radius curves and moments of inertia versus mass are illustrated in Fig.~\ref{MvsR}. (All the calculations presented here were carried
out using {\sc{Mathematica}}~7's default methods.) Specifically, we have chosen the four EOSs from the {\sc{lorene}} library~\cite{LORENE} that are compatible with the Demorest~\emph{et al.}~\cite{Demorestetal} measurement of a $1.97\pm0.04\Msolar$ neutron star (the
nucleonic EOSs APR~\cite{APR}, BBB2~\cite{BBB}, and SLy~\cite{DH} and the hyperonic EOS GNH3~\cite{Glendenning85}; note that BBB2 is only compatible within $2\sigma$). We also consider the nucleonic WFF1
EOS~\cite{WFF}, the most compact EOS considered in~\cite{Readetal}, obtained from the
{\sc{rns}} website~\cite{rns_URL}.\footnote{Note, however, that the high compactness of stars obtained with 
the WFF1 EOS is in part attributable
to the fact that this EOS becomes acausal (i.e., has a sound speed greater than the speed of light) at densities well below the central density of the maximum mass stable stars. The same is true for the EOS we consider that generates the second most compact stars, the APR EOS, which is computed using a variational method,
like the WFF1 EOS. Thus, in our plots, we will mark the points at which stars
constructed with these EOSs start to contain acausal matter.} For more exotic EOSs, we use three of the hadron-quark hybrid EOSs from~\cite{J-MO1} (Hy1, LKR1, and generic) and the strange quark matter EOS constructed in~\cite{J-MO2} using the results of Kurkela, Romatschke, and Vuorinen~\cite{KRV} (KRV1).
Most of these EOSs are $1\sigma$ compatible with the very recent measurement of a $2.01\pm0.04\Msolar$ neutron star by Antoniadis~\emph{et al.}~\cite{Antoniadisetal}.\footnote{But note that the Antoniadis~\emph{et al.}\ measurement relies on some modeling of white dwarf atmospheres, and is thus not as clean as the Demorest~\emph{et al.}~\cite{Demorestetal} measurement considered above.}  However, the GNH3 and LKR1 EOSs are only compatible within
$2\sigma$ (though the GNH3 EOS is close to being $1\sigma$ compatible), while the BBB2 EOS is only compatible within $3\sigma$. (The $3\sigma$ bounds are $1.90$ and $2.18\Msolar$.) 

\begin{figure*}[htb]
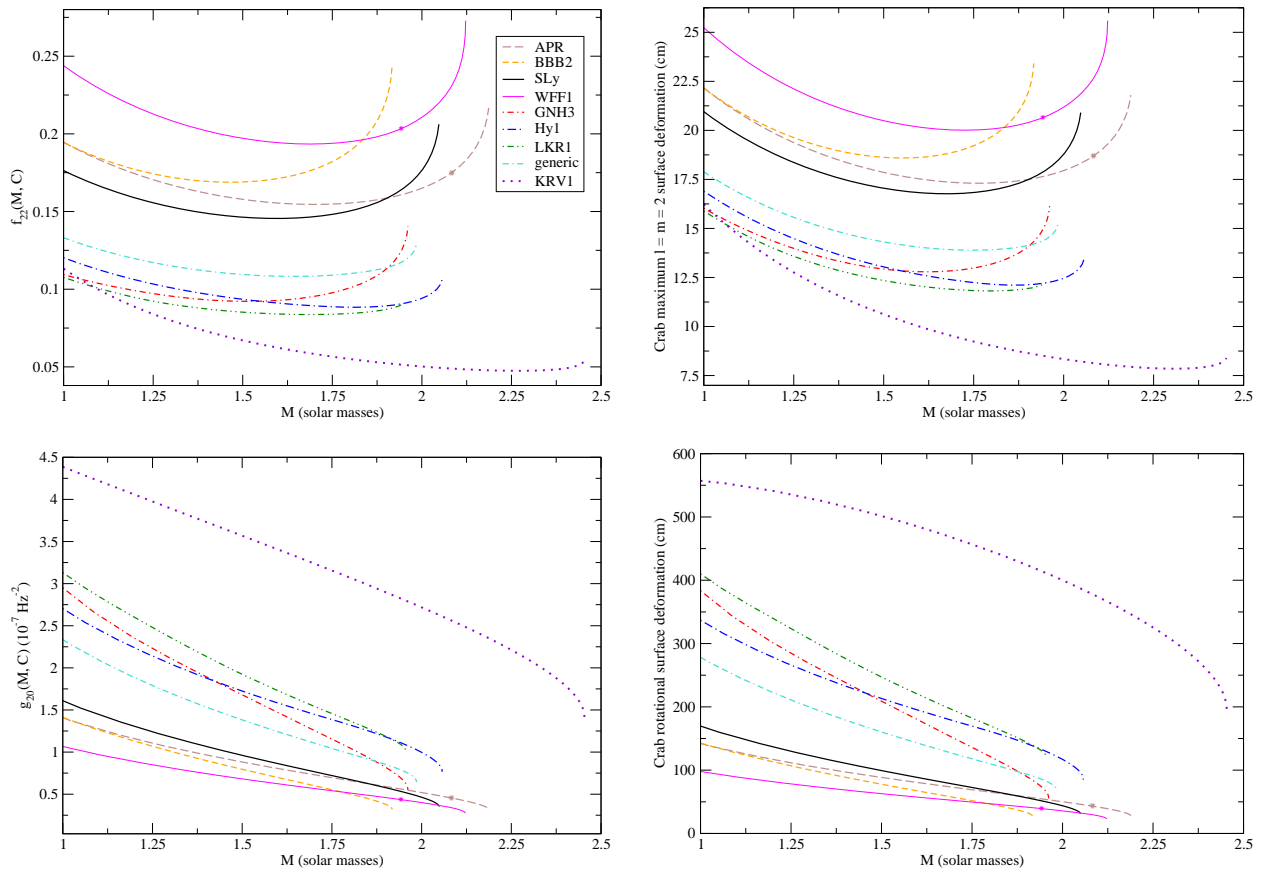

\centering
\subfloat{
\epsfig{file=corr_f_vs_M_WFF.eps,width=8cm,clip=true}
}
\quad
\subfloat{
\epsfig{file=Crab_surf_deformation_WFF.eps,width=8cm,clip=true}
}\\
\subfloat{
\epsfig{file=corr_g20_vs_M_WFF.eps,width=8cm,clip=true}
}
\quad
\subfloat{
\epsfig{file=Crab_surf_deformation_rotation_WFF.eps,width=8cm,clip=true}
}
\caption{\label{corr_func} The $l = m = 2$ \emph{(upper left)} and rotational \emph{(lower left)} conversion functions [$f_{22}(M,\cC)$ and $g_{20}(M,\cC)$] and the maximum $l = m = 2$ \emph{(upper right)} and expected rotational \emph{(lower right)} surface deformations for the Crab pulsar in dimensionful form (i.e., $R\epsilon_{\text{surf, } 22}$ and $R\epsilon_{\text{surf, }20}$), all versus mass and for the various EOSs we consider. 
For the maximum $l = m = 2$ Crab pulsar surface deformations, we use the LIGO upper bound on the fiducial ellipticity of $\epsilon^\text{max}_{\text{fid, } 22} = 10^{-4}$. Also, for the APR and WFF1 EOSs, the asterisks mark the maximum masses that do not contain acausal matter.}
\end{figure*}

We show the conversion factor $f_{22}(M,\cC)$ for these EOSs in Fig.~\ref{corr_func}. Note also that if one considers a
canonical $M = 1.4M_\odot$, $R = 10$~km neutron star, then its compactness is $\sim 0.42$, and $f_{22}(1.4M_\odot, 0.42) \simeq 0.22$. Thus, we see that the fiducial ellipticity generally gives a reasonable
impression of the size of the star's surface deformation. Of course, the actual size of the deformation depends
upon the star's undeformed size: One can give a dimensionful measure of the surface
deformation by multiplying $\epsilon_\surf$ by the
star's radius, $R$ (recalling that the star's equatorial circumference is unchanged by the deformation, to linear order). We show this for the LIGO upper limits on the Crab's quadrupole moment in Fig.~\ref{corr_func}, taking $\epsilon_{\text{fid, } 22} = 10^{-4}$ (around the best upper limit given in~\cite{LIGO_psrs2010}). And while we have used the current upper
limit on the Crab's fiducial ellipticity, one can scale these results linearly
to apply to any other fiducial ellipticity (for the Crab or any other neutron star).

Note, however, that if one considers elastic deformations, then the SLy EOS would not be able to support a quadrupole large enough to create a
fiducial ellipticity anywhere close to $10^{-4}$, unless one assumes a very low-mass star---see Figs.~5 and~6 in~\cite{J-MO2} and Fig.~3 in Horowitz~\cite{Horowitz}. This is simply because the crust is the only solid portion of neutron stars with nucleonic EOSs, like the SLy EOS (and even hyperonic EOSs), and the crustal shear modulus is (relatively) small,
compared with the shear moduli of the higher-density solids that may be present in more exotic EOSs. Thus, while there do not yet exist calculations for the maximum
deformations that could be sustained by the
other hadronic or hyperonic EOSs, they also should not be able to support such large quadrupoles, with standard crustal compositions,
since stars constructed with these EOSs have compactnesses that are roughly the same as those obtained
using the SLy EOS. And even the three hybrid EOSs shown would only be
able to support a quadrupole large enough to create such a large fiducial ellipticity for high masses---see Fig.~9 in~\cite{J-MO2}. However, as shown in Fig.~10 in~\cite{J-MO2}, one could easily obtain such a large
quadrupole with the KRV1 EOS, assuming that the strange matter is in a crystalline state, with a shear modulus given by the Mannarelli, Rajagopal, and Sharma~\cite{MRS} calculation, and that its breaking
strain is $\sim 10^{-1}$, similar to that obtained for the neutron star outer crust in~\cite{HK, HoHe}.\footnote{As discussed in~\cite{J-MO2}, the large breaking strain found for the neutron star outer crust is due to
its high pressure, so it is reasonable to expect that the breaking strain of other materials under similar or
greater pressures to be comparable.} And strong enough internal magnetic fields could sustain such a large
quadrupole for all EOSs. See, in particular, the calculations of magnetic deformations by
Frieben and Rezzolla~\cite{FR} for four of the nucleonic EOSs we consider.

\begin{figure}[htb]
\begin{center}
\epsfig{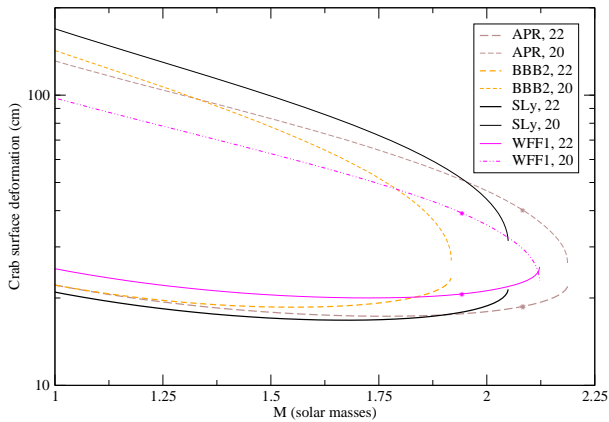}
\end{center}
\caption{\label{surf_comp} The maximum $l = m = 2$ (lower curves) and expected rotational ($l = 2$, $m = 0$; upper curves) dimensionful deformations  (i.e., $R\epsilon_{\text{surf, } 22}$ and $R\epsilon_{\text{surf, } 20}$) of the Crab pulsar versus mass for the hadronic EOSs
we consider, using the LIGO bound of $\epsilon^\text{max}_{\text{fid, } 22} = 10^{-4}$. As before, asterisks denote the maximum masses that do not contain acausal matter for the APR and WFF1 EOSs.}
\end{figure}

Now considering the rotational deformations, we show the conversion factor $g_{20}(M,\cC)$ and the dimensionful surface deformation for the Crab in Fig.~\ref{corr_func}. Again, one can scale the results
for the Crab to any other (slowly rotating) pulsar, using the scaling of the surface deformation with $\nu^2$. 

The behavior of all these quantities is as expected: 
As the star becomes more massive and more compact, one finds that a given $m = 2$ quadrupole moment (i.e., a given fiducial $l = m = 2$ ellipticity) translates into a smaller surface deformation, as one would anticipate, since the star's moment of inertia is increasing. This trend reverses for the highest masses, however, where the relativistic suppressions of the quadrupole due to the increasing compactness
(discussed in Sec.~V of~\cite{J-MO2}) now dominate. Additionally, the more massive,
compact stars do not deform as easily, and this is seen in the mass dependence of the rotational deformation. 
In the strange quark star case, one even sees that the decrease in deformability with
increasing mass is strong enough to dominate the increase in radius.
In general, for a fixed rotation rate and fiducial ellipticity, the $l = m = 2$ surface deformation is the largest and the rotational deformation is the smallest for the most compact stars. This is illustrated using the bounds and expectations for the dimensionful surface deformation of the
Crab pulsar and the most compact EOSs we consider (all nucleonic) in Fig.~\ref{surf_comp}.

\begin{table*}
\begin{tabular}{*{10}{c}}
\hline\hline
 & \multicolumn{2}{c}{Crab} & \multicolumn{2}{c}{Vela} & \multicolumn{2}{c}{J0537$-6910$} & \multicolumn{2}{c}{J1952$+3252$}\\
$\nu$~(Hz) & \multicolumn{2}{c}{$29.8$} & \multicolumn{2}{c}{$11.2$} & \multicolumn{2}{c}{$62.0$} & \multicolumn{2}{c}{$25.3$}\\
$\epsilon_{\text{fid, } 22}^\text{max}$ & \multicolumn{2}{c}{$1.0\times10^{-4}$} & \multicolumn{2}{c}{$1.1\times 10^{-3}$} &  \multicolumn{2}{c}{$1.0\times10^{-4}$} & \multicolumn{2}{c}{$2.3\times 10^{-4}$}\\
\hline
($10^{-5}$) & $\epsilon_{\text{surf, } 22}$ & $\epsilon_{\text{surf, } 20}$ & $\epsilon_{\text{surf, } 22}$ & $\epsilon_{\text{surf, } 20}$ & $\epsilon_{\text{surf, } 22}$ & $\epsilon_{\text{surf, } 20}$ & $\epsilon_{\text{surf, } 22}$  & $\epsilon_{\text{surf, } 20}$\\
(cm) & $R\epsilon_{\text{surf, } 22}$ & $R\epsilon_{\text{surf, } 20}$ & $R\epsilon_{\text{surf, } 22}$ & $R\epsilon_{\text{surf, } 20}$ & $R\epsilon_{\text{surf, } 22}$ & $R\epsilon_{\text{surf, } 20}$ & $R\epsilon_{\text{surf, } 22}$ & $R\epsilon_{\text{surf, } 20}$\\
\hline
\multirow{2}{*}{APR} & $1.6$ & $8.6$ & $18$ & $1.2$ & $1.6$ & $37$ & $3.7$ & $6.2$\\
& $18$ & $97$ & $200$ & $14$ & $18$ & $420$ & $42$ & $70$\\
\hline
\multirow{2}{*}{BBB2} & $1.7$ & $8.0$ & $19$ & $1.1$ & $1.7$ & $35$ & $3.9$ & $5.8$\\
& $19$ & $89$ & $210$ & $13$ & $19$ & $390$ & $43$ & $64$\\
\hline
\multirow{2}{*}{SLy} & $1.5$ & $9.5$ & $16$ & $1.3$ & $1.5$ & $41$ & $3.4$ & $6.8$\\
& $17$ & $110$ & $190$ & $16$ & $17$ & $480$ & $40$ & $80$\\
\hline
\multirow{2}{*}{WFF1} & $2.0$ & $6.6$ & $22$ & $0.93$ & $2.0$ & $29$ & $4.6$ & $4.8$\\
& $21$ & $69$ & $230$ & $9.7$ & $21$ & $300$ & $48$ & $50$\\
\hline
\multirow{2}{*}{GNH3} & $0.93$ & $17$ & $10$ & $2.4$ & $0.93$ & $73$ & $2.1$ & $12$\\
& $13$ & $240$ & $150$ & $34$ & $13$ & $1000$ & $30$ & $170$\\
\hline
\multirow{2}{*}{Hy1} & $0.97$ & $17$ & $11$ & $2.4$ & $0.97$ & $72$ & $2.2$ & $12$\\
& $14$ & $230$ & $150$ & $33$ & $14$ & $1000$ & $31$ & $170$\\
\hline
\multirow{2}{*}{LKR1} & $0.87$ & $19$ & $9.6$ & $2.7$ & $0.87$ & $82$ & $2.0$ & $14$\\
& $13$ & $280$ & $140$ & $39$ & $13$ & $1200$ & $29$ & $200$\\
\hline
\multirow{2}{*}{generic} & $1.1$ & $14$ & $12$ & $1.9$ & $1.1$ & $59$ & $2.6$ & $9.8$\\
& $15$ & $180$ & $160$ & $25$ & $15$ & $780$ & $34$ & $130$\\
\hline
\multirow{2}{*}{KRV1} & $0.73$ & $33$ & $8$ & $4.7$ & $0.73$ & $140$ & $1.7$ & $24$\\
& $11$ & $520$ & $120$ & $73$ & $11$ & $2200$ & $26$ & $370$\\
\hline\hline
\end{tabular}
\caption{\label{comparison_table} The maximum $l = m = 2$ and expected rotational ($l = 2$, $m = 0$) surface deformations 
(both dimensionless and dimensionful) for
the four pulsars for which there are LIGO/Virgo upper bounds below the spin-down limit, for the various EOSs we consider, assuming a $1.4\Msolar$ star. We also give the stars'
rotational frequencies $\nu$, and the upper bounds on fiducial ellipticity from the LIGO/Virgo results,
$\epsilon_{\text{fid, } 22}^\text{max}$. The values for $\nu$ are valid for the years $2005$--$2007$, corresponding to the time of the gravitational wave observations. The Vela pulsar observations were later,
but the Vela pulsar's spin-down is relatively slow, so its spin frequency is not affected, to the given accuracy.
Indeed, the spin frequencies given are the same as the current ones to the given accuracy for all pulsars
except for the Crab: The Crab pulsar spins down rather rapidly, so its current spin frequency is $29.7$~Hz (from the Jodrell Bank Crab Pulsar Monthly
Ephemeris~\cite{LPG-S}), which reduces its
rotational surface deformation by $\sim0.7\%$ compared with the values given in the table. All values for the deformations are rounded to two significant figures.
}
\end{table*}

In Table~\ref{comparison_table}, we compare the 
LIGO/Virgo upper bounds on the $l = m = 2$ deformation with the expected rotational deformations for the four neutron stars for which the LIGO/Virgo bounds are lower than indirect limits, and there is a measured spin
period (viz., the Crab and Vela pulsars, J0537$-6910$, and J1952+3252)~\cite{LIGO_psrs2010, LIGO_Vela}.
We do this assuming the fiducial $1.4\Msolar$ neutron star and considering all the EOSs used in this paper. But recall that there is no reason to assume any correlation between the $l = m = 2$
and $l = 2$, $m = 0$ deformations. 
We merely quote the numbers for rotation as a convenient
scale against which to compare the upper bounds on the nonaxisymmetric deformations.

Additionally, note that the last two stars we consider require a moment of inertia larger than the fiducial one (of $10^{45}\text{ g cm}^2$), or a significantly
closer distance than the estimate used in~\cite{LIGO_psrs2010} in order for the LIGO upper bound to beat the spin-down limit. This is not too much of a concern for J0537$-6910$, as one only needs a factor of $\sim 2$ larger moment of inertia, which is easily provided by any
number of EOSs (see, e.g., this paper's Fig.~\ref{MvsR} and Fig.~6 in~\cite{Bejger}). However, for J1952+3252, \cite{LIGO_psrs2010} claims that LIGO likely just misses beating the spin-down limit, even given
the uncertainties in distance and moment of inertia. We have still included this star here, since the requisite
moment of inertia of $\gtrsim 4\times 10^{45}\text{ g cm}^2$ would be easily attainable with the KRV1 EOS for
higher masses.\footnote{In considering the effects of the moment of
inertia on the spin-down limit, be aware of the following typographical errors in~\cite{LIGO_psrs2010}: In that
paper's Eqs.~(1) and (7), $I_{38}$ should be $I_{38}^{1/2}$ and $I_{38}^{-1}$, respectively. The first of these errors is also present in an inline equation in the first paragraph of Sec.~1 of~\cite{LIGO_Crab} and Eq.~(14) of \cite{LIGO_Vela} [though Eq.~(1.2) in~\cite{LIGO_psrs2007} is correct]. The second is also present in the first paragraph of Sec.~3 of~\cite{LIGO_Crab}, though the analogous Eqs.~(6.1) in~\cite{LIGO_psrs2007} and (15) in~\cite{LIGO_Vela} are correct.} Additionally, there is a fair amount of uncertainty in this pulsar's
distance: Using {\sc{h~i}} measurements, Verbiest~\emph{et al.}~\cite{Verbiestetal} give distance bounds
of $3\pm 2$~kpc, so on the small side, this is significantly closer than the distance of $2.5$~kpc used in~\cite{LIGO_psrs2010}, and would be enough to put the LIGO bound under the spin-down limit, even assuming the fiducial moment of inertia. (However, note that reducing the distance would reduce
the inferred limit on the fiducial ellipticity from the LIGO results, in addition to increasing the spin-down limit, while increasing the moment of inertia does not affect the limit on the fiducial ellipticity.)
But even these large strange quark star moments of inertia are not quite large enough to make the LIGO limit  lie below a reasonable spin-down limit for J1913+1011, the pulsar with the next closest LIGO limit to its fiducial spin-down limit, without assuming that the pulsar
is significantly closer than thought: The LIGO limit for J1913+1011 is $4.9$ times its fiducial spin-down limit, and
the spin-down limit scales as $I_{zz}^{1/2}$---cf.\ Eq.~(1.2) in~\cite{LIGO_psrs2007}. 

Considering the EOSs presented here, we can see that gravitational wave
observations have constrained the large-scale (i.e., $l = 2$) surface deformation on the Crab to be
due to rotation, not to any sort of $l = m= 2$ distortion. [The only exception comes from the near-maximum
mass stars given by the WFF1 EOS, with its high maximum compactness of $0.668$; note that such stars contain a significant
region ($\sim15\%$ by volume) of acausal matter.] This conclusion could \emph{not} have been made solely from electromagnetic observations, since the
star's spin-down only puts a limit on the $l = m = 2$ deformation that is at least $\sim 7$ times greater than the
LIGO limit (and
possibly a factor of a few more, given the uncertainties in the moment of inertia)---see Table~3 in~\cite{LIGO_psrs2010}. This would only be sufficient to constrain the Crab pulsar's $l = m = 2$ surface deformation
to be smaller than its rotational surface deformation for stars with large radii. (Note that some interpretations of X-ray observations of neutron stars~\cite{SLB,SLB2} imply that their radii are $\sim11$--$12$~km, so the EOSs that
give stars with larger radii may be disfavored, though these interpretations are by no means conclusive.) And even the smaller, less direct upper limit obtained by Palomba~\cite{Palomba} by folding in the star's braking index only bounds the deformation to be $\sim 3$ times larger than the LIGO bound, which would be insufficient to
ensure that the $l = m = 2$ surface deformation is smaller than the rotational surface deformation for massive, compact stars, as shown in Fig.~\ref{surf_comp}. (While we have used the best upper limit for the Crab's
fiducial ellipticity given in Table~3 in~\cite{LIGO_psrs2010} in drawing this conclusion, note that improved bounds from LIGO/Virgo are forthcoming~\cite{Gill}, so this is a mild caveat.)

Since the Vela pulsar spins at less than half the frequency of the Crab pulsar, and has an upper bound on its fiducial ellipticity that is an order of magnitude greater, the observations do not bound its $l = m = 2$ surface
deformation to be smaller than its rotational surface deformation, even for the case in which the two deformations are the most similar, the LKR1 EOS and a mass of $\sim 1.8\Msolar$. Conversely, since J0537$-6910$ has
a rotational frequency that is $\sim 2$ times the Crab's and the same bound on its fiducial ellipticity, we know
that its rotational deformation is larger than its $l = m = 2$ deformation for all
masses and EOSs, even for the most massive, compact stars with the WFF1 EOS. (Indeed, the spin is large
enough that this conclusion could have been obtained from the electromagnetic observations of the spin-down, even with the uncertainties in the moment of inertia.) The bounds on J1952+3252's $l = m = 2$ surface deformation compared with its rotational deformation are similar to
the Crab's, though somewhat less good, due to J1952+3252's somewhat smaller spin frequency and the larger upper bound on its fiducial ellipticity. In particular, the bound on J1952+3252's $l = m = 2$ surface
deformation will be larger than its rotational surface deformation for masses slightly larger than $1.4\Msolar$
for the WFF1 EOS, and for all the other nucleonic EOSs for larger masses.

Finally, for a close-to-home
scale against which to compare these surface deformations, the Earth's (dimensionless) rotational surface deformation is $\sim 1.7\times10^{-3}$, and the (dimensionless) $l = m = 2$ surface deformation of the geoid 
(i.e., of a gravitational equipotential near sea level) is $\sim 2.8\times 10^{-6}$. These values were obtained using the value of the flattening of the Earth's reference ellipsoid and the $l = m = 2$ component of the
Earth's gravitational field, both from the WGS84 version of the Earth Gravitational Model EGM2008~\cite{PHKF}, along with the relation between the
flattening and the rotational surface deformation given at the end of Sec.~\ref{e20_comp}. The Crab pulsar's
and J1952+3252's
(dimensionless) rotational surface deformations are thus very similar to the Earth's for the EOSs with larger radii, if the pulsars' masses are close to $1.4\Msolar$. However, all of the bounds on the $l = m = 2$
(dimensionless) surface deformation for the pulsars we consider are larger than the Earth's $l = m = 2$
(dimensionless) surface deformation, though the bounds (electromagnetic or gravitational wave) for many of the faster rotating pulsars are well below this (see Table~1 in~\cite{LIGO_psrs2010}).

\section{Conclusions and outlook}\label{concl}

We have shown that one can convert the bounds on the fiducial ellipticity of neutron stars (from gravitational
wave or electromagnetic observations) into bounds on the shape of the star's surface, in full general relativity, to first
order in the perturbation (and in the slow-rotation limit). We have given the conversion for a variety of EOSs, some purely nucleonic and others containing exotica, including the strange quark star case (with the
requisite changes to the calculation), and compared with the shape of
the star's surface due to rotation. Here we find that the gravitational wave observations have constrained the
Crab pulsar's $l = m = 2$ surface deformation to be smaller than its $l = 2$, $m = 0$ surface deformation
due to rotation, for all the causal EOSs we consider.

For the other three pulsars for which LIGO/Virgo observations have beaten the spin-down limit, within the uncertainties on the star's distance and moment of
inertia, we find that the Vela pulsar's relatively slow rotation means that the bound on its $l = m = 2$
surface deformation is well above its rotational surface deformation, for all the EOSs we consider. On the other hand, for the more quickly rotating
J0537$-6910$, the $l = m = 2$ surface deformation is known to be below the rotational deformation from
the electromagnetic observations of its spin-down. (Again, this holds for all the
EOSs we consider.) For J1952+3252, the bound on the $l = m = 2$ surface
deformation is below the rotational surface deformation for all cases except for heavy stars constructed
with the hadronic EOSs.

Looking into the future, the prospects for improving these bounds are good. Indeed, as mentioned in Sec.~6 of~\cite{LIGO_Vela}, Virgo+ science run 4 (VSR4) was expected to be sensitive to fiducial ellipticities of a $\text{few} \times 10^{-4}$
for the Vela pulsar and $\sim 10^{-5}$ for the Crab pulsar and J1952+3252 if it achieved its sensitivity goal, and preliminary results using data from this run indeed give improved bounds for the Crab and Vela pulsars~\cite{Gill}. These improvements would bound the $l = m = 2$ surface deformation of the Vela pulsar to be below its rotational surface deformation for EOSs with larger radii, and place the bounds for the Crab pulsar and J1952+3252 well
below their rotational surface deformation for all masses and EOSs. (Of course, in the most optimistic scenario, one would obtain a measurement of the $l = m = 2$ surface deformation instead of a bound.)
There are also several other pulsars for which the Enhanced LIGO
and Virgo+ observations already made could be able to beat the spindown limit (as discussed in Sec.~6 of~\cite{LIGO_psrs2010}).

Advanced LIGO
and Virgo are expected to have sensitivities more than 10 times better than the initial interferometers
for the pulsars considered in~\cite{LIGO_psrs2010}, which would allow them to further improve the bounds on the
pulsars already considered and
to place bounds below the spin-down limit for many more pulsars (at least $47$, based on
the results in Sec.~2.2 of Pitkin~\cite{Pitkin}). And looking even further into the future, Pitkin~\cite{Pitkin} predicts that
the Einstein Telescope will place bounds below the spin-down limit for hundreds of pulsars.
Additionally, as shown in Fig.~3 of Pitkin~\cite{Pitkin}, the majority of the stars for which these detectors are expected to beat the
spin-down limit have (relatively) small spin frequencies, so
constraining the $l = m = 2$ surface deformation to be smaller than the
expected $l = 2$, $m = 0$ rotational deformation gives another (vaguely physical) benchmark for searches, below the spin-down limit. However,  we must stress once again that there is no expected relation between the
two deformations.

On the theoretical side, it would be interesting to compute the rotational surface deformation of more
rapidly rotating stars using either {\sc{lorene}}~\cite{LORENE} or {\sc{rns}}~\cite{rns_URL} 
and compare with the slow-rotation predictions given here. Such a calculation, or the more involved calculations likely needed to obtain the bounds on the $l = m = 2$ surface deformations of more rapidly rotating stars from bounds on their gravitational wave emission, could be of interest in interpreting results from next-generation detectors, which are
expected to be able to beat the spin-down limit (by a bit) for one or two pulsars with spin frequencies of $\sim500$~Hz (see Fig.~3 in Pitkin~\cite{Pitkin}). However, these frequencies are still low enough (compared to the Kepler
frequency) that our slow-rotation approximation is likely still reasonably accurate (within tens of
percent). (See the values for the discrepancy in the quadrupole moment in matching a Hartle-Thorne slow-rotation solution to an exact numerical solution as a function of rotation parameter in Table~6 of Berti~\emph{et al.}~\cite{BWMB}, and the values for the Kepler frequency in, e.g., Fig.~2 of Lo and Lin~\cite{LoLin}.)

Likely more important would be the inclusion of the magnetic field in the calculation of the conversion from the $m = 2$
quadrupole moment to the surface deformation: As discussed in the Appendix, it is possible that
small departures from a force-free configuration at the star's surface could affect the conversion for standard
pulsar magnetic fields. And even if the magnetic field configuration is force-free to better accuracy than we need, the order-of-magnitude estimates of its effects on the computation given in Eq.~\eqref{B-eff} suggest that for magnetar-level surface magnetic fields of $\sim 10^{15}$~G, and the fiducial ellipticities of around $10^{-6}$ associated with
internal fields of this magnitude (as given in, e.g., Sec.~7 of Frieben and Rezzolla~\cite{FR}), the effects of the magnetic field could be large enough to affect the results we have given for the shape. However, it is worth
noting that the internal field could be significantly larger than the surface field (as discussed in, e.g.,
Corsi and Owen~\cite{CO}), in which case the corrections would be significantly smaller. Additionally, if the
magnetic field is in the twisted torus configuration with a large toroidal component, then, as Ciolfi and Rezzolla~\cite{CR} have very recently shown, the fiducial ellipticities can be considerably larger, $\sim 10^{-4}$, at least for a polytropic equation of state, which would also significantly reduce the correction.

\acknowledgments

We thank B.~Br{\"u}gmann, B.~J.~Owen, and the anonymous referee for useful comments. This work was supported by the DFG SFB/Transregio 7.

\appendix*

\section{Corrections due to non-perfect fluid contributions}

Since we have obtained our results for the $l = m = 2$ shape deformation using expressions that were derived assuming an unmagnetized perfect fluid star, we need to check that they
are still valid in the cases in which we are interested, where the star is perturbed not by the tidal field of a companion (as in
Damour and Nagar~\cite{DN}), but by some internal stresses. (Since we assume that we can treat these
internal stresses as a first-order perturbation, we do not need to worry about their effects on the rotational deformation, since the two perturbations will be independent, to the order we are computing.)  It turns out that we can indeed apply these expressions with no
changes, if one assumes that the stresses at the star's surface (times $R^2$ or $1/\rho_-$) are much smaller than $H_0(R)$, which we shall see is the case in the
majority of situations of interest. Explicitly, one needs to make this assumption in obtaining the expression for the star's perturbed surface using its enthalpy perturbation [see Eq.~(26) in Damour and Nagar~\cite{DN}~], and also in writing the metric function
$K(R)$ in terms of $H_0(R)$ and $H_0'(R)$ [see the discussion around Eq.~\eqref{dRRH0K}].

In the first case, the expression for the enthalpy used to obtain the position of
the perturbed surface is obtained from stress-energy conservation. The relevant equation including the
stress terms is given in Eq.~(A2) of Ipser~\cite{Ipser}, and with slightly different notation in Eq.~(35) of~\cite{J-MO2}. In the second case, Ipser gives the relation between $K$ and $H_0$ including the stress terms in Eq.~(39a).

We now consider the extent to which realistic surface stresses affect these results. Looking at the second case
first, the star's magnetic field produces a stress of $\sim B^2/8\pi$, so the fractional corrections from these stresses to the relation between $K$ and $H_0$ will be on the order of
\<\label{B-eff}
\begin{split}
&\sim\frac{B^2R^2}{H_0(R)} = \frac{B^2M^3R^2}{\pi Q_{22}\bar{H}_0(R)}\\
&\simeq 10^{-9}\left(\frac{M}{1.4\Msolar}\right)^3\left(\frac{R}{10\text{ km}}\right)^2\left(\frac{B}{10^{12}\text{ G}}\right)^2
\left(\frac{10^{-4}}{\epsilon_{\text{fid, }22}}\right).
\end{split}
\?
[We have omitted the
dependence on the star's compactness from $\bar{H}_0(R)$ in the final expressions here and later, though it will further
suppress these corrections for higher-mass stars.]

Considering 
the corrections to the stress-energy
conservation equation used to locate the star's surface, we have fractional corrections on the order of
$RJ^aF_{ab}/\rho_s H_0(R)$, since the addition to stress-energy conservation from the electromagnetic field is $\nabla^bT^\mathrm{EM}_{ab} = J^b F_{ab}$, where $J^a$ and $F_{ab}$ denote the (4-)current density and Faraday tensor,
respectively. [Compare Eq.~(35) in~\cite{J-MO2}; the magnetic field term would appear on the right-hand side of that equation, and the factor of $R$ comes
from the factor of $\nabla_a Y_{lm}$ that multiplies everything else in that
expression.]

We expect $J^aF_{ab}$ (which gives the Lorentz force density) to be small, compared to $\rho_s/R$, since the magnetic field in the magnetosphere is expected to be
close to force-free (as is discussed, e.g., at the end of Sec.~2.2 in~\cite{CFGP}). If one takes the
magnetosphere to be \emph{exactly} force-free (as is frequently done in models of magnetically deformed neutron stars), then $J^aF_{ab}$ is zero. However,  in the situations we consider, $J^aF_{ab}$ could be
of the same order as the $H_0$ term, since in this case, the Lorentz force
could help balance the gravitational perturbation, instead of the pressure perturbation alone balancing it, which is what occurs in the absence of additional stresses. But treating this properly is a task for far more
detailed neutron star modeling (analogous to the issue with surface currents discussed in Sec.~II~C of Corsi and Owen~\cite{CO}), so we simply take the magnetic field configuration to be exactly force-free here.

For the Crab pulsar, with a surface field of $\sim 4\times 10^{12}$~G, one would have to consider a fiducial ellipticity of at most $\sim 10^{-11}$ for the first of these corrections to approach the percent level. This is well below the
predicted sensitivity of even the proposed Einstein Telescope (see, e.g., Fig.~3 in Pitkin~\cite{Pitkin}), and  
around
the minimum ellipticity quoted in~\cite{LIGO_Crab} as being produced by an internal field of the same magnitude as the
Crab's external field. (The Frieben and Rezzolla~\cite{FR} fits predict an ellipticity of $\sim 10^{-9}$ for an internal field of that strength.) The other
three pulsars considered in Table~\ref{comparison_table} have external magnetic fields of similar magnitudes~\cite{MHTH}, so this analysis holds for them, as well.

On the other hand, if we consider a solid strange quark star, the surface shear stresses could be quite large, if the gap parameter does not decrease too much as one approaches the star's surface and the high breaking strain assumed for the high-pressure interior
continues to the surface---see the discussion in Sec.~IV~C of~\cite{J-MO2}. However, it seems likely that
the stresses will be considerably smaller in any sort of reasonable case, since one expects both the gap
parameter and breaking strain to decrease as the pressure decreases. And if one assumes that the surface is not strained much more than the star's average strain, then the surface stress cannot be too large
in the cases of interest.

Specifically, considering the KRV1 strange quark EOS, we have ratios of (in order
of magnitude, and putting in the $\sim 15.5$~km radius appropriate for a $1.4\Msolar$ star with this EOS,
in addition to the associated values of the Mannarelli, Rajagopal, and Sharma~\cite{MRS} shear modulus)
\<
\begin{split}
&\sim\frac{\mu_-\sigma_-R^2}{H_0(R)} = \frac{M^3\mu_-\sigma_-R^2}{\pi Q_{22}\bar{H}_0(R)}\\
&\simeq 10^{-3}\left(\frac{M}{1.4\Msolar}\right)^3\left(\frac{R}{15.5\text{ km}}\right)^2\left(\frac{\mu_-}{10^{33}\text{ erg cm}^{-3}}\right)\\
&\quad\times\left(\frac{\sigma_-}{\epsilon_{\text{fid, }22}}\right),
\end{split}
\?
and
\<
\begin{split}
&\sim\frac{\mu_-\sigma_-}{H_0(R)\rho_-} =  \frac{M^3\mu_-\sigma_-}{\pi Q_{22}\bar{H}_0(R)\rho_-}\\
&\simeq 10^{-1}\left(\frac{M}{1.4\Msolar}\right)^3\left(\frac{\mu_-/\rho_-}{10^{-2}}\right)
\left(\frac{\sigma_-}{\epsilon_{\text{fid, }22}}\right),
\end{split}
\?
where $\mu_-\sigma_-$ is the shear stress just inside the star's surface (written as a product of shear modulus $\mu_-$
and shear strain $\sigma_-$). (These come, as before, from the
corrections to the relation between $K$ and $H_0$ and to the enthalpy expression used to locate the star's surface.)

While it is possible that
$\sigma_-$ could be as large as $\sim 10^{-1}$, around the breaking strain for the neutron star outer crust obtained in~\cite{HK, HoHe} (and thus orders of magnitude larger than $\epsilon_{\text{fid, }22}$ for the stars we are
considering), this high breaking strain comes from high pressure, so the breaking strain at the
surface, where the pressure goes to zero, will likely be much less. Moreover, if one assumes that the star's surface is strained at about
the same level as the star's average strain, then we will have $\sigma_- \simeq \epsilon_{\text{fid, }22}$ (cf.\ Fig.~10 in~\cite{J-MO2}, remembering that one can
linearly scale the maximum quadrupoles given there for a uniform strain of $10^{-1}$ to any uniform strain).
Thus, we feel comfortable quoting values for
the strange quark star case, since the caveats seem fairly mild in reasonable situations.

\bibliography{paper2}

\end{document}